# Finite Element Formalism for Micromagnetism

H. Szambolics [a], L. D.Buda-Prejbeanu [b], J. C. Toussaint [a,b] and O. Fruchart [a]

[a] Institut Néel, CNRS-INPG-UJF, 25 rue des Martyrs, 38042 Grenoble cédex 9, France

[b] Laboratoire SPINTEC, CEA-CNRS-INPG-UJF, 17 rue des Martyrs, 38054 Grenoble cédex 9, France

Research paper

**Purpose -** The aim of this work is to present the details of the finite element approach we developed for solving the Landau-Lifschitz-Gilbert equations in order to be able to treat problems involving complex geometries.

**Design/methodology/approach -** There are several possibilities to solve the complex Landau-Lifschitz-Gilbert equations numerically. Our method is based on a Galerkin-type finite element approach. We start with the dynamic Landau-Lifschitz-Gilbert equations, the associated boundary condition and the constraint on the magnetization norm. We derive the weak form required by the finite element method. This weak form is afterwards integrated on the domain of calculus.

**Findings -** We compared the results obtained with our finite element approach with the ones obtained by a finite difference method. The results being in very good agreement, we can state that our approach is well adapted for 2D micromagnetic systems.

**Research limitations/implications -** The future work implies the generalization of our method to 3D systems. To optimize our approach spatial transformations for the treatment of the magnetostatic problem will be implemented.

**Originality/value -** The paper presents a special way of solving the Landau-Lifschitz-Gilbert equations. The time integration a backward Euler method has been used, the time derivative being calculated as a function of the solutions at times *n* and *n+1*. The presence of the constraint on the magnetization norm induced a special two-step procedure for the calculation of the magnetization at instant n+1.

## 1. Introduction

Thanks to high-resolution fabrication techniques (lithography, patterning, self-assembly), submicron magnetic systems are now routinely fabricated with different materials and precisely controlled sizes and shapes. The shape can be dots, with various forms, rings, wires and rods, antidotes, etc. (Li et al., 2004, Jubert et al., 2001). To understand in detail what happens inside such a magnetic system, with both ultimate time and space resolution, accurate experimental studies and micromagnetic modeling must be combined.

Nowadays most of the micromagnetic softwares are based on the finite difference (FD) approximation, meaning that the magnetic body is divided into regular orthorhombic cells. From a numerical point of view, the implementation of these algorithms is straightforward and due to the periodic discretization, the use of Fast Fourier Transforms is possible, thus the computation time is significantly reduced (Toussaint et al., 2002). Furthermore, specific integration schemes were developed to integrate the Landau-Lifschitz-Gilbert (LLG) equation describing the magnetization dynamics (Brown, 1963) which conserve implicitly the magnitude of magnetization. Unfortunately, the algorithms based on FD are intrinsically affected by the roughness of the grid at surfaces. Thus only the systems bounded exclusively by planar surfaces parallel to some axes of the grid, can be in principle reliably computed (García-Cervera et al., 2003).

To overcome these numerical difficulties, an alternative is the finite element method (FEM) (Braess, 2001), which uses an irregular mesh. The characteristic of FEM is that it is based on the projection of the micromagnetic equations on so called test functions. Thus the mathematical background is more complex than in the case of the FD approach, where the physical quantities are estimated locally. Up to now the micromagnetic calculations using irregular mesh, presented by physicists are not based on a projective form of the LLG equation. We show here the steps we followed when building up our FEM approach and the results obtained for two 2D magnetic test cases.

## 2. Weak form for micromagnetism

The principle of micromagnetics is to approximate the magnetization distribution inside a magnetic system with a continuous medium (Brown, 1963). This requires that the variations of the magnetization vector $\mathbf{M(r)} = M_s\, \mathbf{m(r)}$ occur on a length scale large enough to approximate the direction angles of neighboring atomic spins with a continuous function. The spontaneous magnetization $M_s$ denotes the mean magnetic moment per unit volume and is assumed to be constant, only the orientation of the magnetization vector $\mathbf{m(r)}$ may change in time and space.

The magnetization distribution corresponding to an equilibrium state of the ferromagnet is obtained by minimizing the total free energy $E_{tot}$ of the system with respect to **m(r)**. In the continuous medium approximation, $E_{tot}$ is the sum of exchange interactions, magnetocrystalline energy, the Zeeman contribution due to applied field and the dipolar interactions. In the simplest case of uniaxial magnetocrystalline anisotropy of 2$^{nd}$ order, it can be written as:

$$E_{tot}[\mathbf{m}] = \int_\Omega A_{ex}[\nabla \mathbf{m}]^2 d\Omega + \int_\Omega K_1\left[1-(\mathbf{u_K}\cdot\mathbf{m})^2\right]d\Omega \\ - \int_\Omega \mu_0 M_S \mathbf{m}\cdot\mathbf{H_{app}} d\Omega - \int_\Omega \frac{1}{2}\mu_0 M_S \mathbf{m}\cdot\mathbf{H_m} d\Omega \quad (1)$$

This integral expression depends on the material parameters which are $A_{ex}$ the exchange constant, $K_1$ and $\mathbf{u_K}$ for the anisotropy and $M_s$. $\mathbf{H_m}$ is the magnetostatic field, solution of Maxwell's equations. It coincides with the demagnetizing field produced inside the ferromagnet by the magnetization itself. While the terms describing the anisotropy, the exchange interactions, and the Zeeman coupling are local terms, the demagnetizing field depends on the magnetization distribution over the entire material, thus it remains the most difficult term to compute.

The method adopted here to relax the magnetic configuration towards an equilibrium state consists in integrating the LLG dynamic equations (Brown, 1963):

$$\frac{\partial \mathbf{m}}{\partial t} = -\mu_0 \gamma (\mathbf{m}\times\mathbf{H_{eff}}) + \alpha\mu_0\gamma\left(\mathbf{m}\times\frac{\partial \mathbf{m}}{\partial t}\right) \quad (2)$$

Here $\mu_0$ is the gyromagnetic factor, $\alpha$ is the damping constant and $\mathbf{H_{eff}}$ is the effective field obtained by variational derivation of the total energy $E_{tot}$ with respect to **m(r)**. According to (1) the effective field is the sum of four fields: exchange field $\mathbf{H_{ex}}$, anisotropy field $\mathbf{H_{ani}}$, magnetostatic field $\mathbf{H_m}$ and applied field $\mathbf{H_{app}}$. The LLG equation respects implicitly the condition imposed on the norm of the magnetization:

$$1 - \mathbf{m}^2 = 0 \quad (3)$$

Solving these equations by using FEM, as explained below, means deriving their weak form and integrating this by using the most suitable integration method (Braess, 2001).

*A. Weak form of the magnetostatic equations*

The evaluation of $\mathbf{H_m}$ is the most difficult issue because of its long-range character. Let us consider two of Maxwell's equations in magnetostatics:

$$\begin{aligned}\nabla \cdot \mathbf{B} &= 0 \\ \nabla \times \mathbf{H_m} &= 0\end{aligned} \quad (4)$$

the magnetization being related to the magnetic induction $\mathbf{B}$ and to the demagnetizing field $\mathbf{H_m}$ by:

$$\mathbf{B} = \mu_0(\mathbf{H_m} + \mathbf{M}) \quad (5)$$

When working with the magnetic vector potential, the starting point is the solenoidal nature of the $\mathbf{B}$ vector. This means that it is possible to write it as the curl of a vectorial quantity, called magnetic vector potential $\mathbf{A}$, and the magnetostatic field becomes:

$$\mathbf{H_m} = \frac{1}{\mu_0}\mathbf{B} - \mathbf{M} = \frac{1}{\mu_0}\nabla \times \mathbf{A} - \mathbf{M} \quad (6)$$

The potential $\mathbf{A}$ is assumed to vanish at infinity:

$$\mathbf{A}(|\mathbf{r}| \to \infty) \to 0 \quad (7)$$

To derive its weak form, the magnetostatic equation (6) is multiplied by a vector weighting function $\mathbf{v}$ and then integrated over the whole space $\Omega$:

$$\int_\Omega \mathbf{v} \cdot \nabla \times \mathbf{H_m}\, d\Omega = 0 \quad (8)$$

By using:

$$\mathbf{div}(\mathbf{v} \times \mathbf{H_m}) = \mathbf{H_m} \cdot \nabla \times \mathbf{v} - \mathbf{v} \cdot \nabla \times \mathbf{H_m} \qquad (9)$$

and the continuity condition of the tangential component of $\mathbf{H_m}$ at free surfaces, the derivation orders are being equilibrated and the weak formulation reads as:

$$\int_\Omega \nabla \times \mathbf{v} \cdot \nabla \times \mathbf{A}\, d\Omega = \mu_0 \int_\Omega \mathbf{M} \cdot \nabla \times \mathbf{v}\, d\Omega \qquad (10)$$

Due to the invariance of the studied system along the Oz direction only the z component of the vector potential $\mathbf{A}$ must be considered, and since $\mathbf{v}=(0, 0, v)$ the final form of the weak formulation is:

$$\int \partial_x v \left(\partial_x A_z + \mu_0 M_y\right) + \partial_y v \left(\partial_y A_z - \mu_0 M_x\right) = 0 \qquad (11)$$

For the treatment of the condition at infinity (7) a spatial transformation is used. This converts the infinite exterior that must be considered for this problem into a finite domain, so the "open boundary problem" becomes a "closed boundary problem" (Brunotte et al., 1992). The 2D system is thus modified in order to apply the transformation: the upper and the lower semi-infinite regions are converted in two finite domains bounded by straight lines: $-Y_\infty < Y \leq -Y_0$ and $Y_0 \leq Y < Y_\infty$, as depicted in figure 1:

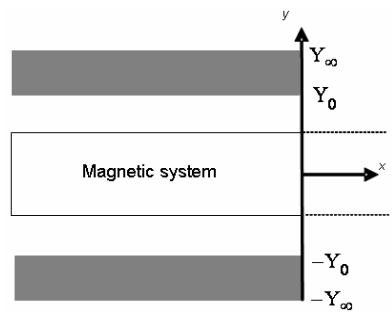

Fig. 1. 2D test case. The upper and the lower semi-infinite regions are replaced by two finite domains coloured in grey.

The capital letters refer to the coordinates in the transformed domains, while the coordinates in small letters indicate the real space.

Since **H<sub>m</sub>** is expected to decay exponentially at long distances away from the magnetic system, a natural choice for the transformation to be used is:

$$y = T(Y) = \text{sgn}(Y) \cdot \left( Y_0 - \frac{T_0}{2\pi} Log \left| \frac{Y_\infty - |Y|}{Y_\infty - Y_0} \right| \right) \qquad (12)$$

## B. Weak form of the micromagnetic equations

To obtain the weak form of the LLG equations we project these onto vector test functions **w**:

$$\int_{\Omega_m} \mathbf{w} \cdot \left[ \frac{\partial \mathbf{m}}{\partial t} - \alpha \mu_0 \gamma \left( \mathbf{m} \times \frac{\partial \mathbf{m}}{\partial t} \right) \right] d\Omega = -\mu_0 \gamma \int_{\Omega_m} \mathbf{w} \cdot \left[ \mathbf{m} \times (\mathbf{H}_{ex} + \mathbf{H}_{ani} + \mathbf{H}_m + \mathbf{H}_{app}) \right] d\Omega \qquad (13)$$

In this case, the integration is done only on the magnetic volume $\Omega_m$.

Only the term that contains the exchange field, $\mathbf{H}_{ex} = 2 A_{ex} \Delta \mathbf{m} / (\mu_0 M_S)$, needs to be transformed. For this term we need to equilibrate the orders of derivatives of the unknowns and of the test functions. In Cartesian coordinates this integrand can be rewritten as follows:

$$(\mathbf{m} \times \mathbf{H}_{ex}) \cdot \mathbf{w} = \frac{2 A_{ex}}{\mu_0 M_S} \sum_{l=\{x,y,z\}} \Delta m_l [(\mathbf{m} \times \mathbf{e}_l) \cdot \mathbf{w}] \qquad (14)$$

By using the divergence theorem, after integration on the magnetic volume, the weak form for the exchange term is obtained:

$$\int_{\Omega_m} (\mathbf{m} \times \Delta \mathbf{m}) \cdot \mathbf{w} \, d\Omega = \sum_{l=\{x,y,z\}} \oiint_S \nabla m_l \cdot \mathbf{n} \cdot [(\mathbf{m} \times \mathbf{e}_l) \cdot \mathbf{w}] dS \\ - \sum_{l=\{x,y,z\}} \int_{\Omega_m} \nabla m_l \cdot \nabla [(\mathbf{m} \times \mathbf{e}_l) \cdot \mathbf{w}] d\Omega \qquad (15)$$

The micromagnetic theory imposes that no exchange torque acts at the free surface. This implies a Neumann condition $\nabla m_l \cdot \mathbf{n} = 0$ on $S$, known as the Brown condition, so the surface integral from (15) vanishes.

Finally after considering also the constraint (3), equation (13) transforms into the weak form:

$$\int_{\Omega_m} \mathbf{w} \cdot \left( \frac{\partial \mathbf{m}}{\partial t} - \alpha \mu_0 \gamma \, \mathbf{m} \times \frac{\partial \mathbf{m}}{\partial t} \right) d\Omega = \gamma \frac{2 A_{ex}}{M_S} \sum_{l=\{x,y,z\}} \int_{\Omega_m} \left( \mathbf{m} \times \frac{\partial \mathbf{m}}{\partial x_l} \right) \cdot \frac{\partial \mathbf{w}}{\partial x_l} d\Omega \\ - \gamma \mu_0 \int_{\Omega_m} \mathbf{w} \cdot \left[ \mathbf{m} \times \left( \mathbf{H}_{ani} + \mathbf{H}_{app} + \mathbf{H}_m \right) \right] d\Omega + \int_{\Omega_m} \mu \left(1 - \mathbf{m}^2\right) d\Omega + \int_{\Omega_m} \lambda \, \mathbf{w} \cdot \frac{\partial \left(1 - \mathbf{m}^2\right)}{\partial \mathbf{m}} d\Omega \quad (16)$$

Here $\lambda$ is the Lagrange multiplier used for the treatment of the constraint and $\mu$ is the corresponding test function.

Another alternative weak form has been proposed by Alouges, where the test functions $\mathbf{w}$ at each mesh node belong to the tangential plane to $\mathbf{m}$ (Alouges et al., 2006). The comparison between the two methods is in progress.

## C. Finite Element Discretization

The expression (16) corresponds to an ideal weak form for the LLG equations. In this ideal case, after finite element discretization, the Lagrange multiplier $\lambda$ and the vector field $\mathbf{m}$ are written as a sum of basis functions $\{\varphi_i\}$ weighted by a set of fitting coefficients. The test functions $\mu$ and $\mathbf{w}$ are calculated using the same functions $\{\varphi_i\}$. We have used as basis functions $2^{nd}$ order Lagrange polynomials.

In practice the constraint on the magnetization norm is applied only at the mesh nodes and only the magnetization and the test functions $\mathbf{w}$ are interpolated:

$$\mathbf{m} = \sum_{j=1}^{N} \varphi_j \mathbf{m}_j = \sum_{j=1}^{N} \varphi_j \sum_{q=\{x,y,z\}} m_{j,q} \mathbf{e_q} \qquad (17)$$
$$\mathbf{w} = \varphi_i \mathbf{e_p}$$

where $N$ is the number of nodes and $p \in \{x, y, z\}$.

The temporal scheme is based on the Euler's backward time integration method where the time derivative is estimated as a finite difference of the solutions at times $n$ and $n+1$:

$$\frac{\partial \mathbf{m}}{\partial t} = \frac{\mathbf{m}^{n+1} - \mathbf{m}^n}{\Delta t} \qquad (18)$$

and the exchange term is also calculated at time $n+1$. The other terms in the right hand member of the equation (16) are evaluated at time $n$.

Due to the constraint (3) a two step procedure has to be implemented. An estimation of the magnetization $\tilde{\mathbf{m}}^{n+1}$ at time step $n+1$ is firstly determined without taking into account the influence of the constraint. Then, a correction $\boldsymbol{\delta}\mathbf{m}$ due to the constraint on the magnetization norm is calculated. The magnetization at time $n+1$ is finally obtained by summing up the two contributions:

$$\mathbf{m}^{n+1} = \tilde{\mathbf{m}}^{n+1} + \boldsymbol{\delta}\mathbf{m} \qquad (19)$$

Inserting the interpolated expressions (17) and the time derivative of the magnetization (18) into the weak form the following matrix equation is obtained, from which the solution $\tilde{\mathbf{m}}^{n+1}$ is calculated:

$$\left( M_{ij,pq} + D_{ij,pq} + \Delta t\ K_{ij,pq} \right) \tilde{m}_{j,q}^{n+1} = \left( M_{ij,pq} + D_{ij,pq} \right) m_{j,q}^n \qquad (20)$$

where

$$M_{ij,pq} = \int_{\Omega_m} \varphi_i \varphi_j\ \delta_{pq} d\Omega$$
$$D_{ij,pq} = \int_{\Omega_m} \varphi_i \varphi_j\ \alpha \left(\mathbf{e_p} \times \mathbf{e_q}\right) \cdot \mathbf{m}^n d\Omega \qquad (21)$$
$$K_{ij,pq} = \gamma \frac{2 A_{ex}}{M_S} \sum_{l=x,y,z} \int_{\Omega_m} \frac{\partial \varphi_i}{\partial x_l} \frac{\partial \varphi_j}{\partial x_l} \left[\left(\mathbf{e_p} \times \mathbf{e_q}\right) \cdot \mathbf{m}^n\right] d\Omega$$

with implicit summation over $j \in \{1,...,N\}$ and $p \in \{x, y, z\}$.

By introducing the constraint term in the equation the following set of equations is obtained:

$$\begin{cases} \left(M_{ij,pq} + D_{ij,pq} + \Delta t\ K_{ij,pq}\right) \delta m_{j,q} + H^T_{ip,j} \lambda_j = 0 \\ H_{i,jq}\ \delta m_{j,q} = G_i\left(\tilde{\mathbf{m}}^{n+1}\right) \end{cases} \quad (22)$$

where $\lambda_j$ represents a Lagrange multiplier which expresses the presence of the constraint and $H_{ij,q}$ is the associated Jacobian matrix:

$$\begin{aligned} H_{ij,q} &= 2\delta_{ij}\ \tilde{m}^{n+1}_{j,q} \\ G_i &= 1 - \sum_q \left(\tilde{m}^{n+1}_{i,q}\right)^2 \end{aligned} \quad (23)$$

To simplify the expressions used up to now the equation set (22) is rewritten by using the matrix notation:

$$\begin{cases} \left(M + D + \Delta t\ K\right) \boldsymbol{\delta m} + H^T \boldsymbol{\Lambda} = 0 \\ H\ \boldsymbol{\delta m} = G\left(\tilde{\mathbf{m}}^{n+1}\right) \end{cases} \quad (24)$$

The general solution of (22) may be written as follows:

$$\boldsymbol{\delta m} = Nul\ \mathbf{u} + \mathbf{m_d} \quad (25)$$

where *Nul* is the matrix that collects all the vectors of Ker(H), and therefore satisfies $H\ Nul = 0$.

The Lagrange multiplier can be eliminated from the first equation in (22) by multiplying it with $Nul^T$. One finally obtains:

$$\begin{cases} Nul^T\left(M + D + \Delta t\ K\right)\left(Nul\ \mathbf{u} + \mathbf{m_d}\right) = 0 \\ H\mathbf{m_d} = G\left(\tilde{\mathbf{m}}^{n+1}\right) \end{cases} \quad (26)$$

The method for solving (26) consists in determining firstly $\mathbf{m_d}$ and then $\mathbf{u}$. From the reduced singular value decomposition (SVD) of $H$

$$H = U_r S_r V_r^T \quad (27)$$

where $S_r$ is a positive definite matrix, the expression for $\mathbf{m_d}$ is obtained:

$$\mathbf{m_d} = V_r S_r^{-1} U_r^T G(\tilde{\mathbf{m}}^{n+1}) \tag{28}$$

It is now possible to determine $\mathbf{u}$ from (26) by replacing $\mathbf{m_d}$ with (28):

$$\mathbf{u} = -K_{eff}^{-1} Nul^T (M + D + \Delta t\, K)\, \mathbf{m_d} \tag{29}$$

where $K_{eff}$ denotes:

$$K_{eff} = Nul^T (M + D + \Delta t\, K) Nul \tag{30}$$

## 3. Test cases

We present here an application of our finite element approach to magnetic thin films with a perpendicular anisotropy of moderate strength. To this category belong FePd alloys, Co/Pt multilayers or $Co(10\bar{1}0)$. The equilibrium magnetization configuration of such systems consists of a periodic modulation of the perpendicular component of the magnetization leading to parallel stripe domains (Toussaint et al., 2002). This kind of configuration is well adapted to 2D micromagnetic simulations since the magnetization is nearly invariant along the stripes' direction (Oz axis) and is periodic in the other in-plane direction (Ox axis). Due to these features, the simulations are done for only one period of the system of length L=200 nm and thickness h=40 nm. A schematic representation of the model system is given in figure 2:

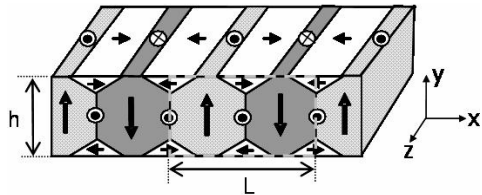

Fig. 2. Schematic representation of the stripe structure in a thin film.

To test our approach the relaxation process to equilibrium calculated by FEM is compared with the one obtained by a finite difference approach implemented in the GL_FFT software (by J.C. Toussaint, © Lab. Louis Néel).

As initial magnetization configuration a sinusoidal profile has been chosen. The magnetization distribution is depicted in the figure 3.a):

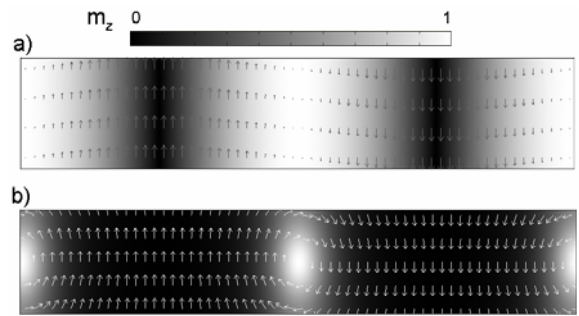

Fig. 3.a) The initial magnetic configuration. b) The equilibrium magnetization distribution calculated by FEM. For both a) and b) the $m_x$ and $m_y$ magnetization components are represented by arrows and the $m_z$ component by a grey scale. Material parameters: $A_{ex}=2·10^{-11}$ J/m, $\mu_0 M_s=1$T and $K_1=10^5$ J/m$^3$.

By monitoring the time evolution of the total energy, depicted in figure 4, we verify if the time integration scheme describes a dissipation process towards equilibrium. A small energy gap around 1% is observed at equilibrium between the FD and FEM calculations. For such physical systems our FEM approach is thus validated. The residual gap can be attributed to the different ways to evaluate the total energy: FD uses local estimations of the magnetization vector and the effective field, whereas in

FEM the energy expression (1) is applied to the magnetization field interpolated on each element.

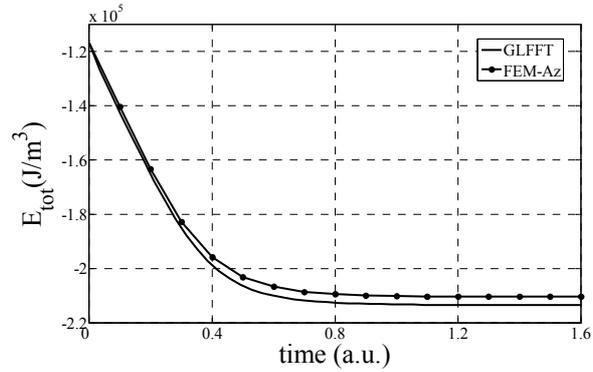

Fig.4. Time evolution of the total energy density.

We apply now our approach to test cases which, in principle, can be dealt only with FEM. Such a system is a thin film with artificial periodic constrictions (Fig. 5.). The material parameters considered for this system are the same as before. The dimension of the geometry is: length L=200 nm and full thickness h=65 nm. The starting magnetization configuration is similar to that considered for the regular geometry, but is phase shifted with 45°. As expected the domain walls drift during the equilibration process and finally they are located on the constrictions. The domain walls are pinned on the constrictions, minimizing the energy of the system. This result proves the feasibility of our numerical approach in analyzing systems with complex shape.

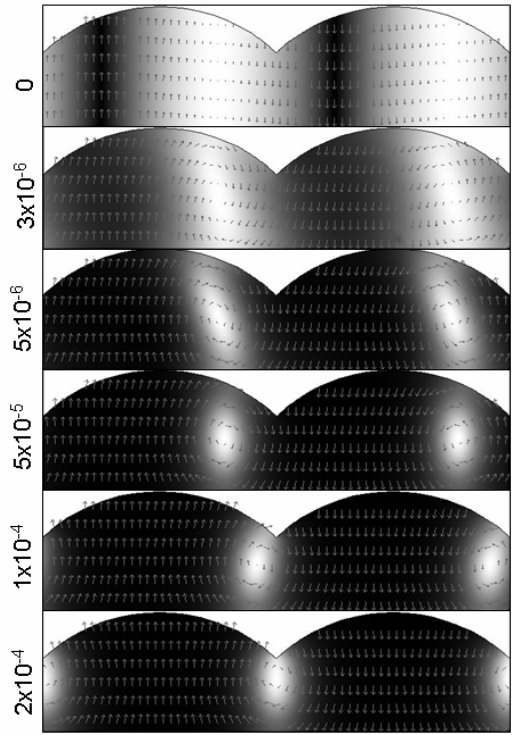

Fig. 5. Magnetization distribution calculated by FEM at several times in arbitrary units. The same grey scale is used as previously.

## 4. Conclusions

The weak form of the LLG equations presented here seems to be well adapted to deal with 2D micromagnetic systems. Its generalization to 3D systems is in progress and requires for magnetostatics a special treatment of the open boundary based on spherical shell transformations (Brunotte et al., 1992).